\documentclass[twocolumn,prb,aps,longbibliography]{revtex4-1}
\usepackage{dcolumn}
\usepackage{amsmath}
\usepackage{amsfonts}
\usepackage{amssymb}
\usepackage{calligra}
\usepackage{bm}
\usepackage{graphicx}
\usepackage{float}
\usepackage{subfigure}
\usepackage{stackengine}
\usepackage{natbib}
\usepackage{hyperref}

\newcommand{\avg}[1]{\left< #1 \right>} % for average
\newcommand{\abs}[1]{| #1 |} % for absolute value
\newcommand{\ket}[1]{| #1 \rangle} % for Dirac bras
\newcommand{\bra}[1]{\langle #1 |} % for Dirac kets

\DeclareMathAlphabet{\mathcalligra}{T1}{calligra}{m}{n}

\begin{document}

\title{Distribution of entanglement Hamiltonian spectrum in free fermion models}
\author{Mohammad Pouranvari}
\affiliation {Department of Physics, Faculty of Basic Sciences,
  University of Mazandaran, P. O. Box 47416-95447, Babolsar, Iran}

% \pacs{}
\date{\today}

\begin{abstract}
  We studied numerically the distribution of the entanglement
  Hamiltonian eigenvalues in two one-dimensional free fermion models
  and the typical three-dimensional Anderson model. We showed
  numerically that this distribution depends on the phase of the
  system: In the delocalized phase it is centered around very small
  values and in the localized phase, picks of the distribution goes to
  larger values. We therefore, based on the distribution of
  entanglement Hamiltonian eigenvalues, explain the behavior of the
  entanglement entropy in different phases. In addition we propose the
  smallest magnitude entanglement Hamiltonian eigenvalue as a
  characterization of phase and phase transition point (although it
  does not locate the phase transition point very sharply), and we
  verify it in the mentioned models.
  
\end{abstract}

\maketitle

\section{Introduction}
Concept of Entanglement was firstly employed in the field of quantum
information science\cite{PhysRevLett.81.5932,PhysRevLett.67.661,
  RevModPhys.74.145,PhysRevLett.80.869, doi:10.1119/1.1463744,
  Kane1998} as a resource of information, now it is used in the
condensed matter physics\cite{RevModPhys.81.865,LAFLORENCIE20161,
  PhysRevLett.90.227902}. Since it measures indirectly the correlation
among the system, people use it as a non-local phase
characterization. In particular, this concept is useful in the
Anderson phase transition between delocalized and localized
phases\cite{PhysRev.109.1492}. In the localized phase, where state of
the system is localized, we expect lower correlation compare to the
delocalized phase with extended states. In the same manner, we expect
lower entanglement in the localized phase compare to the delocalized
phase\cite{PhysRevLett.99.126801,PhysRevB.89.115104}.

There are several measures of entanglement to quantify
it\cite{PhysRevLett.78.2275, RevModPhys.81.865}, among which the
entanglement entropy (EE) attracted more attention. It has been used
vastly before, specially when the system is in a pure ground state
where EE is a reliable quantity to measure entanglement (there are
other useful measures for a mixed highly excited state\cite{Alba_2009,
  2018arXiv180804381L}). EE is the von Neumann entropy of the reduced
density matrix for a chosen subsystem in a bipartite system. This
partitioning can also be made in the momentum
space\cite{PhysRevLett.110.046806} rather than in real space, or it
can be even a random partition\cite{rosz2019entanglement,
  PhysRevB.91.220101}. There are several examples of using EE for
detecting phase transition point, we mention some of them below. In
Ref. [\onlinecite{Osterloh2002}], connection between quantum
information and a quantum critical point is explained and entanglement
is used as a scaling quantity near phase transition
point. Entanglement properties of an interaction spin-$1/2$ model is
studied in Ref. [\onlinecite{PhysRevA.69.022107}] which shows a
diverging behavior at the critical point. Relation between
discontinuity of Hamiltonian energy and entanglement is studied in
Ref. [\onlinecite{PhysRevLett.93.250404}]. Beside the ground state
application of the EE, we can also mention works that utilized the
entanglement notion for a highly excited
state\cite{PhysRevA.97.042115, PhysRevLett.110.091602, Ares_2014,
  Alba_2009, Caputa2015} and also out of equilibrium
states\cite{2019arXiv190200025G, 2019arXiv190404270P}, although there
are many other applications\cite{PhysRevA.68.042330,
  PhysRevLett.90.227902, PhysRevA.69.054101}.

Beside the EE, people also use the entanglement spectrum, which is
spectrum of reduced density matrix, to distinguish different
phases. Li and Haldane used the low lying entanglement spectrum to
identify the topological order\cite{PhysRevLett.101.010504}. Also
degeneracy of the entanglement spectrum was shown to be the property
of the Haldane phase of $S=1$\cite{PhysRevB.81.064439}. Moreover,
distribution of the reduced density spectrum is obtained in the
scaling regime of critical point which depends only on the central
charge\cite{PhysRevA.78.032329}. There are also other
applications\cite{PhysRevLett.105.115501,PhysRevLett.108.196402,
  PhysRevLett.105.080501, PhysRevB.83.245134, PhysRevLett.105.116805,
  PhysRevLett.109.237208, 0295-5075-119-5-57003, PhysRevB.95.115122,
  PhysRevLett.104.130502}. Furthermore, some attempts were made to use
eigenstate of the entanglement Hamiltonian as a quantity that carries
useful physical information\cite{PhysRevB.89.115104,
  PhysRevB.88.075123, PhysRevB.92.245134}.

In this report, we focus on the entanglement Hamiltonian; we show that
entanglement Hamiltonian spectrum (EHS), i.e. the eigenvalues of the
entanglement Hamiltonian, have useful physics information regarding
the delocalized-localized phase transition for a free fermion model in
the ground-state. First, the probability distribution of the EHS is
noticeably different in delocalized and localized phases. In localized
phase, distribution is narrowed around large eigenvalues, and as we go
toward delocalized phase, it becomes narrowed around smaller
eigenvalues. Second, we derive a phase characterization from
distribution of EHS: The smallest magnitude eigenvalue has distinct
behavior in delocalized and localized phases. To verify our ideas, we
use one-dimensional free fermion models and also the typical
three-dimensional Anderson model, both have delocalized-localized
phase transition as we change the disorder strength in the system.

Structure of the paper is as follows: In section \ref{model} we
briefly explain the models we use in this paper, and also methods of
calculating the entanglement Hamiltonian eigenvalues. Distrubution of
the EHS is studied in section \ref{dist} for delocalized and localized
phases, to show their distinguishable behaviour. In section \ref{eps}
we introduce a new characterization for the delocalized-localized
phase transition. A summary is given in the section \ref{sum}.

\section{Models and Method}\label{model}
We start by introducing the main concepts regarding the
entanglement. We consider a system with a pure many-body eigenstate
$\ket{\Psi}$ at zero temperature. Then, density matrix will be
$\rho= \ket{\Psi} \bra{\Psi}$. We the bipartite system into two
subsystems $A$ and $B$. For each subsystem the reduced density matrix
is obtained by tracing over degrees of freedom of the other subsystem:
$\rho^{A}=tr_{B} (\rho)$. Block von Neumann entanglement entropy
between the two subsystems is
$EE=-tr(\rho^{A}\ln{\rho^{A}})=-tr(\rho^{B}\ln{\rho^{B}})$. For a
single Slater-determinant ground state, the reduced density matrix of
each subsystem can be written as:
\begin{equation} \label{rho}
  \rho^{A}=\frac{1}{Z} e^{-H^{A}},
\end{equation}
where $H^{A}$ is the free-fermion \emph{entanglement} Hamiltonian
($Z$ is determined by $tr \rho^{A}=1$):
\begin{equation}
H^{A} = \sum_{ij} h_{ij}^{A}  c_{i} ^{\dagger} c_{j},
\end{equation}
where $c^{\dagger}_i(c_i)$ is the creation (annihilation) operator for
the site $i$ in the second quantization representation.

To calculate entanglement energies $\epsilon$'s, i.e. the eigen-values
of the $h^{A}$ matrix we use correlation function\cite{Klich_2006}. We
diagonalize the correlation matrix of a subsystem, say $A$
\begin{equation}
C_{i,j}=\avg{c_{i} ^{\dagger} c_{j}},
\end{equation}
(where $i$ and $j$ go from $1$ to $N_A$) and find its eigen-values
$\{ \zeta \}$. Eigen-values of the correlation matrix and those of the
entanglement Hamiltonian are related as:
\begin{equation}\label{zeta2eps}
\zeta_i=\frac{1}{1+e^{\epsilon_i}},
\end{equation}
and EE will be given as:
\begin{equation}\label{EE}
\text{EE}=-\sum_{i=1} ^{N_A} [\zeta_i \ln(\zeta_i)+(1-\zeta_i) \ln(1-\zeta_i)],
\end{equation}

We use three models to study our criteria of Anderson phase
transition. First model we use, is power-law random banded matrix
model (PRBM)\cite{PhysRevE.54.3221} which is a $1d$
long range hopping model with the following Hamiltonian:
\begin{equation}\label{Hhij}
  H = \sum_{i,j=1}^N h_{ij}c^{\dagger}_i c_j,
\end{equation}
(where $N$ is the system size) matrix elements $h_{ij}$ are independent random numbers, distributed by a Gaussian
distribution function that has with zero mean and the following
variance (when we use periodic boundary condition):
\begin{equation}
  \avg {\abs{h_{ij}}^2} = \left[{1+\left(\frac{\sin{\pi
            (i-j)/N}}{b \pi /N}\right)^{2a}}\right]^{-1},
\end{equation}
To calculate the entanglement properties, we divide the $1d$ system into two equal subsystems. Subsystem $A$ is from site $1$ to site $N/2$, and the rest is the subsystem $B$. The system is delocalized for $a<1$; at the phase transition point
$a=1$, it undergoes Anderson localization transition to localized
states for $a>1$. This phase transition happens regardless of $b$, and
in our calculation we set $b=1$.

Another model is power-law random bond Anderson model
(PRBA)\cite{PhysRevB.69.165117} which is a $1d$ model with the following
Hamiltonian:
\begin{equation}\label{Hhij}
H = \sum_{i,j=1}^N h_{ij}c^{\dagger}_i c_j,
\end{equation}
where on-site energies $h_{ii}$ are zero, and the hopping amplitudes
are:
\begin{equation}
h_{ij}= w_{ij}/|i-j|^{a}
\end{equation}
where $w$'s are independent uniformly random numbers distributed between $-1$ and
$1$. To calculate the entanglement properties, we divide the $1d$ system into two equal subsystems. Subsystem $A$ is from site $1$ to site $N/2$, and the rest is the subsystem $B$. There is a phase transition at $a=1$ between delocalized state
($a<1$) and localized state ($a>1$). 

Another model we use is the Anderson model in three-dimensional $3d$
space, with the following Hamiltonian:

\begin{equation}\label{oh}
  H=t\sum_{<i,j>} (c_i^{\dagger} c_{j}+c_{j}^{\dagger} c_i)+
  \sum_{i}\epsilon_i c_i^{\dagger} c_i,
\end{equation}
where $<>$ indicates nearest neighbor hopping only. Hopping amplitudes
are constant $t=-1$, and on-site energy $\epsilon_i$ are independent
random numbers distributed with Gaussian distribution with mean zero
and variance $w$. Anderson phase transition happens at $w_c$, below
which state are delocalized and above which states are localizes.
$w_c \approx 6$\cite{1367-2630-16-1-015012}. To calculate the entanglement properties, we divide the $3d$ system into two equal subsystems. The entire system has $N \times N \times N$ sites. Subsystem $A$ is from site $1$ to site $N \times N \times N/2$, and the rest is the subsystem $B$. We use open boundary condition.

\section{Distribution of Entanglement Hamiltonian spectrum}\label{dist}
We already know the behaviour of the EE in a free fermion model with
delocalized-localized phase transition\cite{PhysRevB.89.115104}. In
delocalized phase, eigenstate of the system is extended and we expect
larger EE compare to the localized phase. Thus, by looking at the
behavior of the EE as we change the disorder in the system, we can
distinguish different phases. On the other hand, we can look at the
behavior of the EE as we increase system size, $N$ with a fixed value
of disorder. In delocalized phase EE increases with system size and
violate the area law\cite{RevModPhys.82.277, PhysRevB.93.205120,
  Alba_2019}, while it saturates to a fixed value in localized
phase. In addition, we can look at EE from EHS point of
view. Eq. (\ref{EE}) tells us that among all eigenvalues of
correlation function $\{\zeta\}$, those $\zeta$'s close to $1/2$ have
bigger share in the EE; or in terms of EHS (see Eq. (\ref{zeta2eps})),
those $\epsilon$'s close to zero are the most effective spectrum in
the EE, and as we move away from zero, $\epsilon$'s become less
effective. So the distribution of EHS is informative. 

\begin{figure}
  \centering
  \begin{subfigure}{}%
    \includegraphics[width=0.435\textwidth]{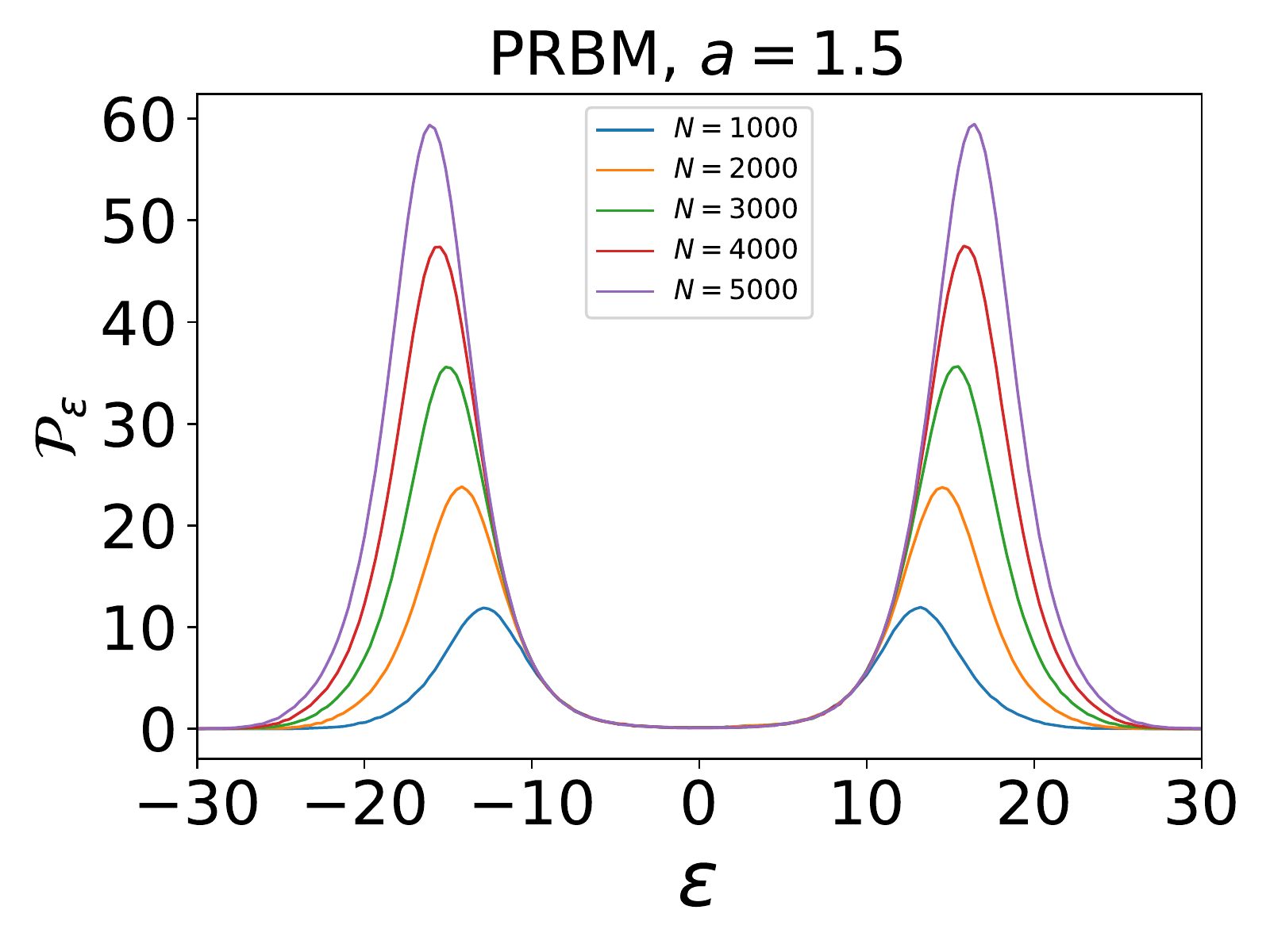}
  \end{subfigure}%
  ~%
  \begin{subfigure}{}%
    \includegraphics[width=0.435\textwidth]{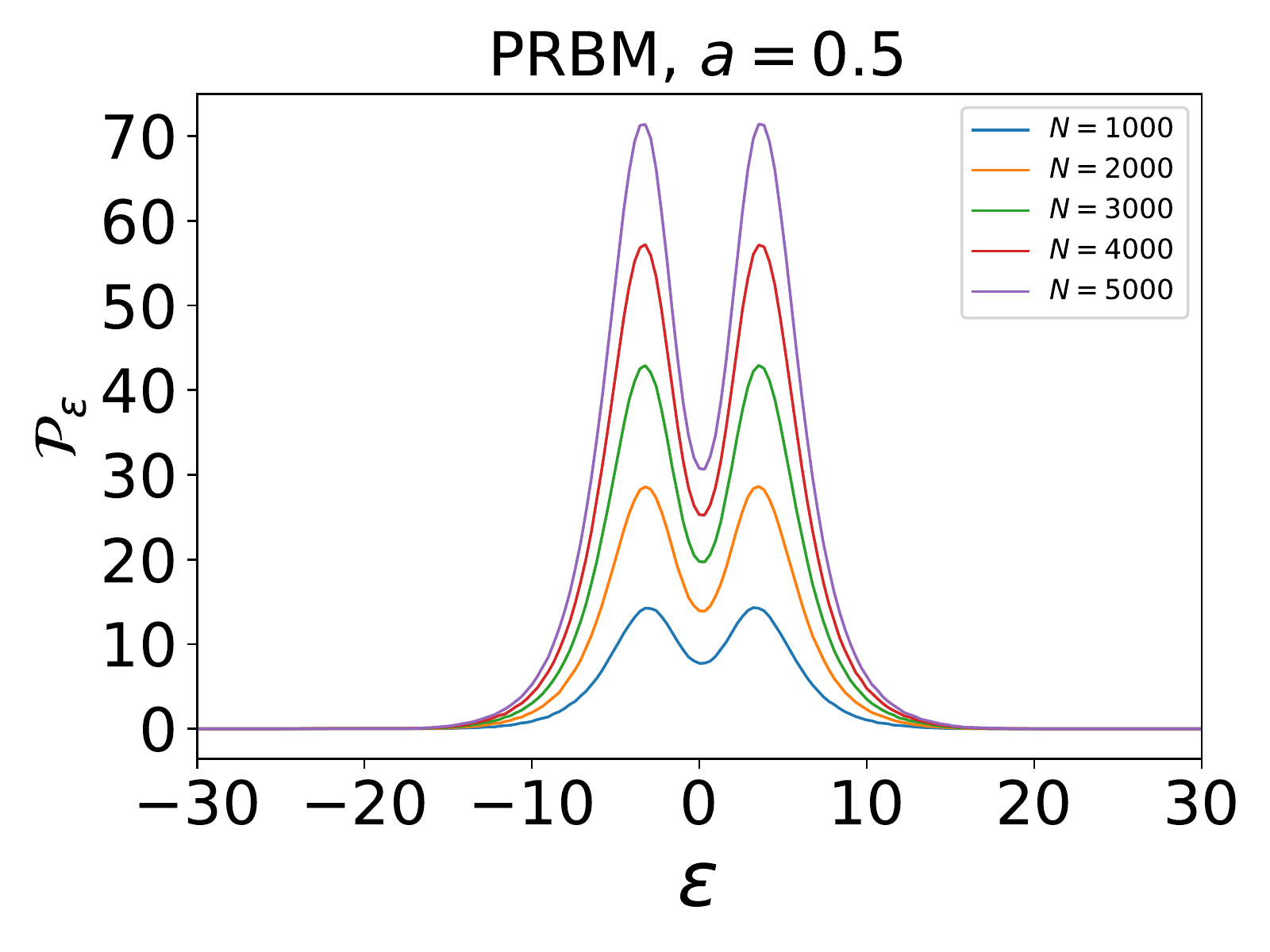}
  \end{subfigure}
  \caption{(color online) probability distribution of the entanglement
    Hamiltonian spectrum for PRBM model in localized (upper panel) and
    delocalized (lower panel) phase. Distribution is plotted for
    different system sizes $N=1000, 2000, 3000, 4000, 5000$ from
    bottom to top. Number of samples ranges between $20000$ for small
    system sizes and $1000$ for large system
    sizes.  \label{fig:disEHS_PRBM}}
\end{figure}

\begin{figure}
  \centering
  \begin{subfigure}{}%
    \includegraphics[width=0.435\textwidth]{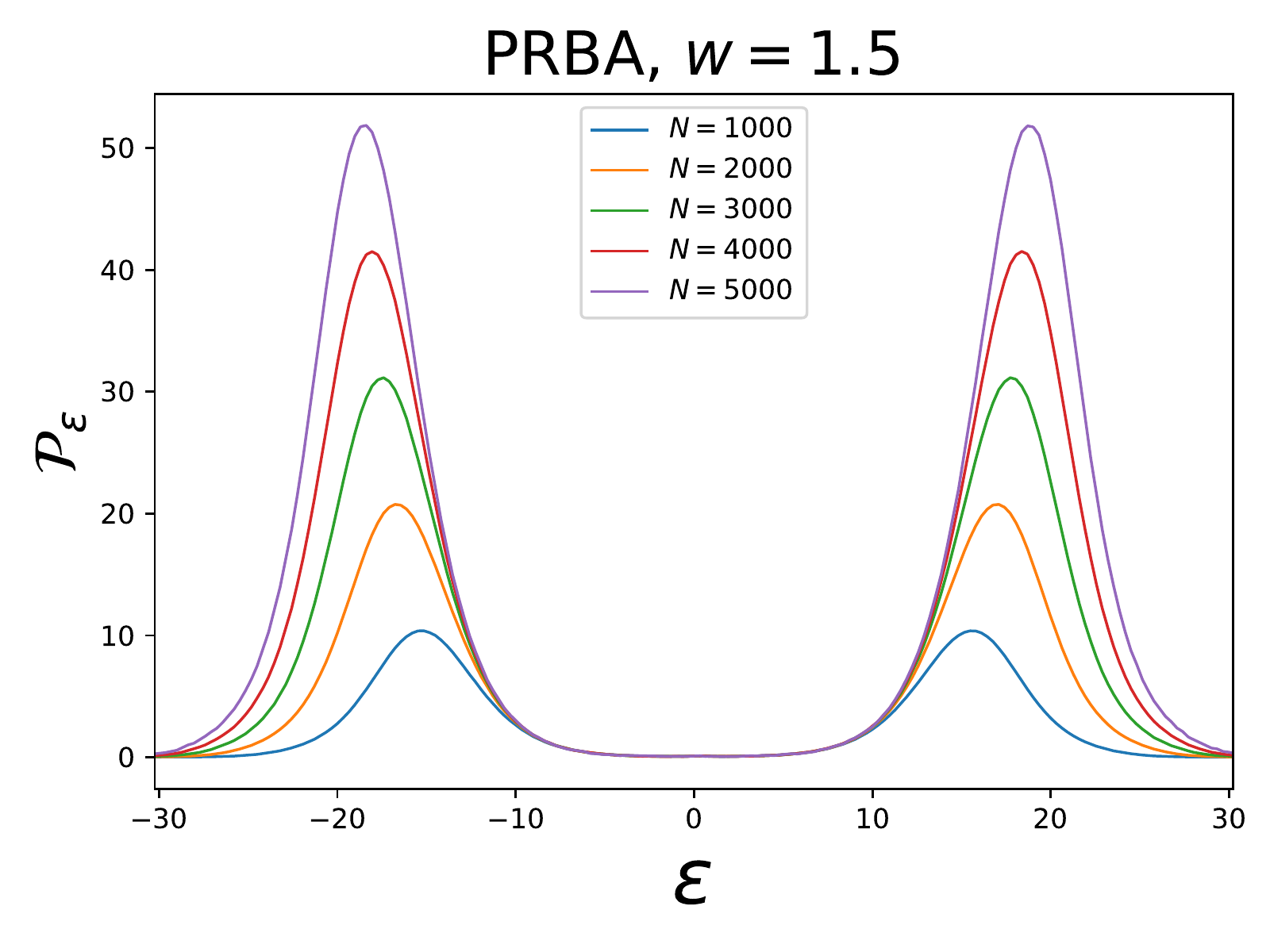}
  \end{subfigure}%
  ~%
  \begin{subfigure}{}%
    \includegraphics[width=0.435\textwidth]{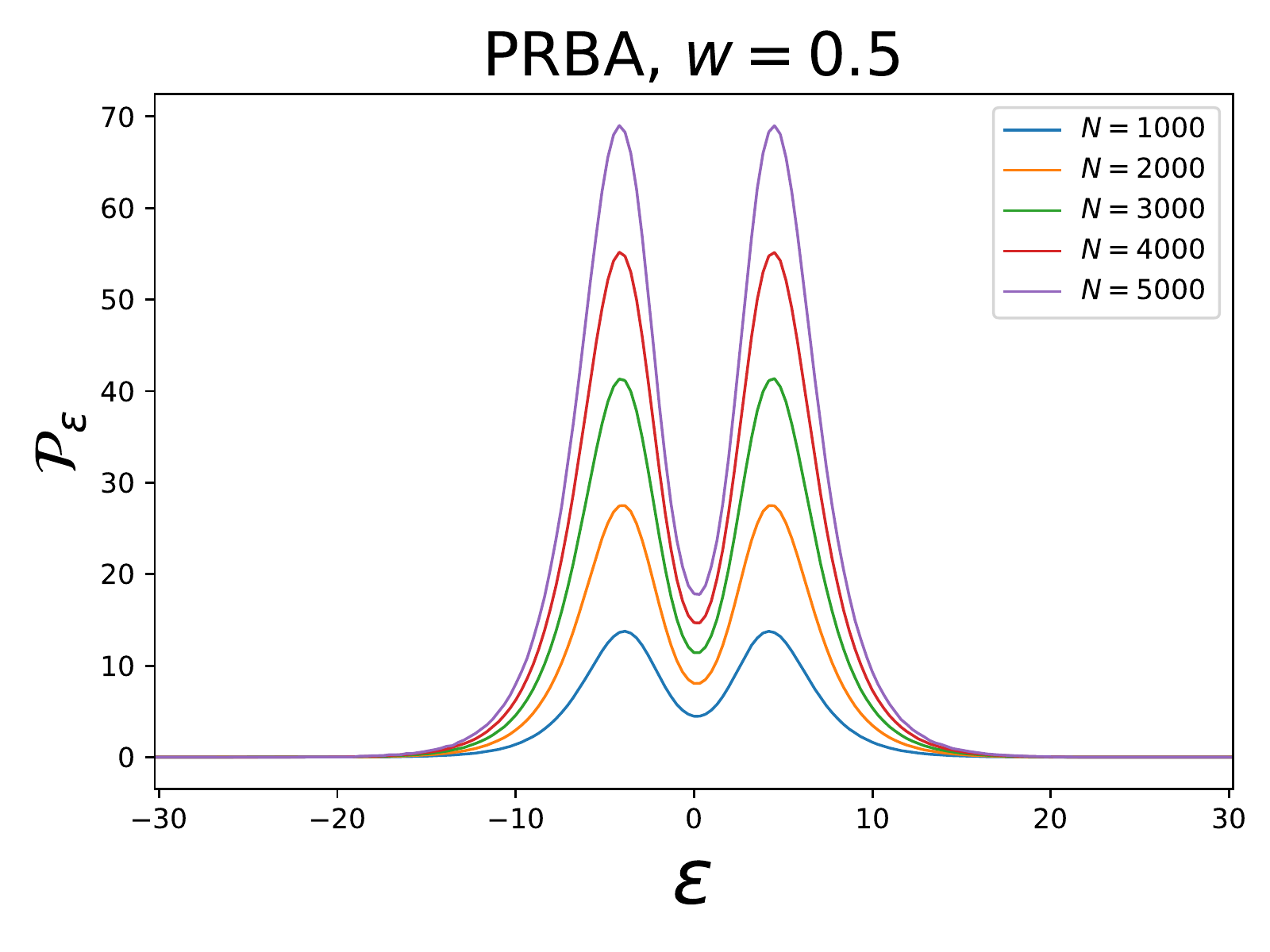}
  \end{subfigure}
  \caption{(color online) probability distribution of the entanglement
    Hamiltonian spectrum for PRBA model in localized (upper panel) and
    delocalized (lower panel) phase. Distribution is plotted for
    different system sizes $N=1000, 2000, 3000, 4000, 5000$ from
    bottom to top. Number of samples ranges between $20000$ for small
    system sizes and $1000$ for large system
    sizes.  \label{fig:disEHS_PRBA}}
\end{figure}

\begin{figure}
  \centering
  \begin{subfigure}{}%
    \includegraphics[width=0.435\textwidth]{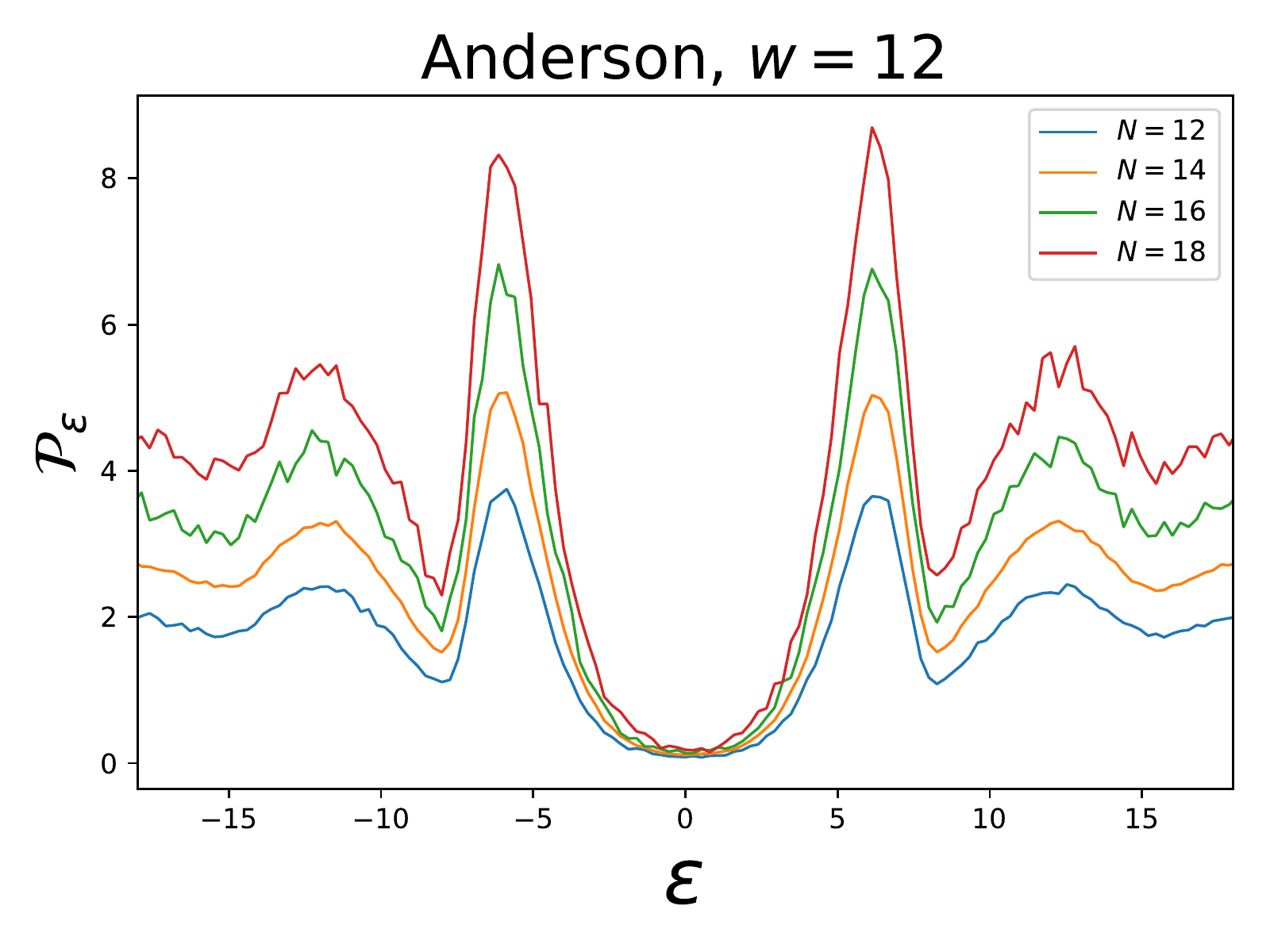}
  \end{subfigure}%
  ~%
  \begin{subfigure}{}%
    \includegraphics[width=0.435\textwidth]{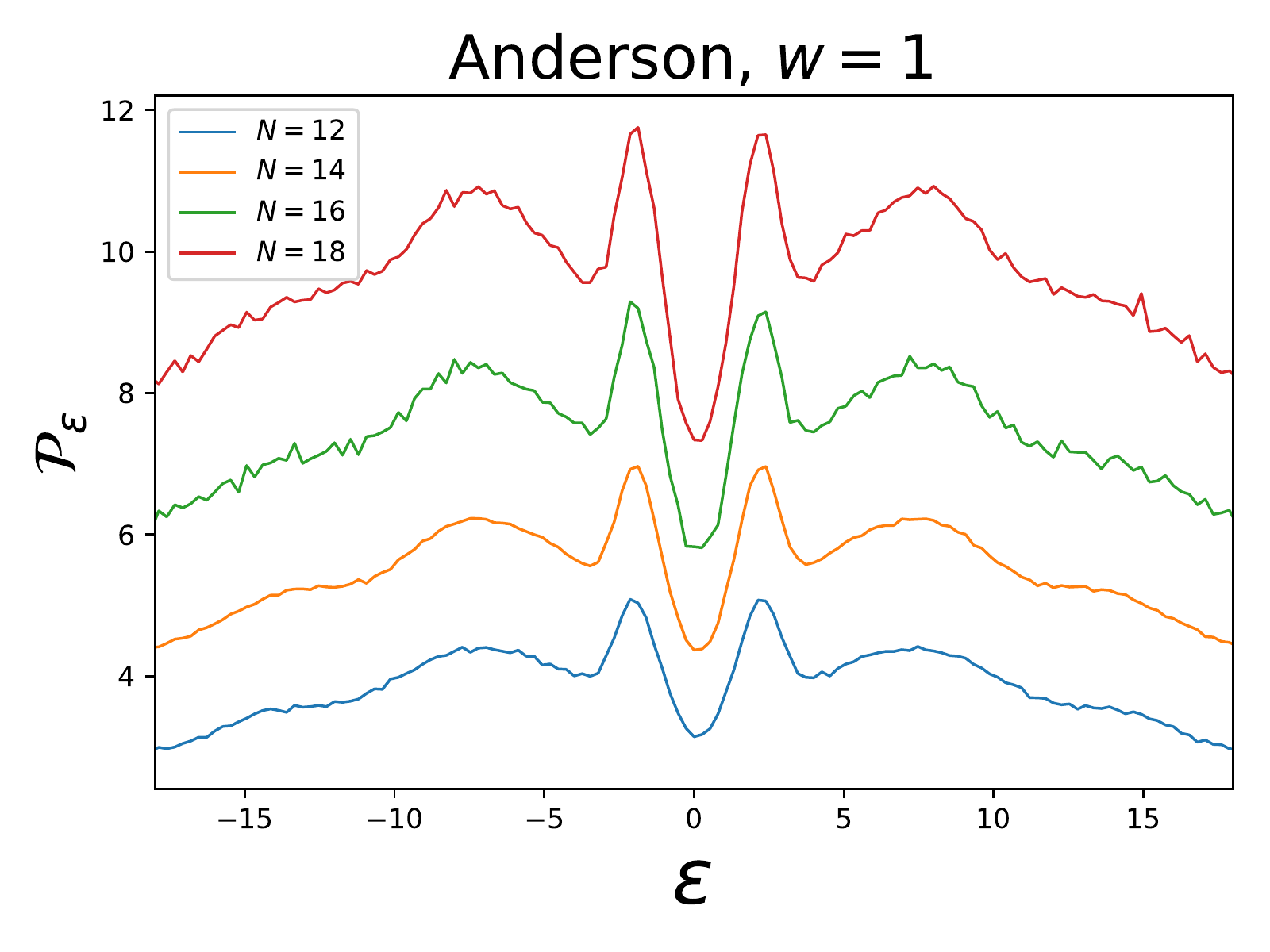}
  \end{subfigure}
  \caption{(color online) probability distribution of the entanglement
    Hamiltonian spectrum for Anderson $3d$ model in localized (upper
    panel) and delocalized (lower panel) phase. Distribution is plotted
    for different linear system sizes $N=12, 14, 16, 18$ from bottom
    to top, where system size is $N \times N \times N$. Number of
    samples ranges between $2000$ for small system sizes and $200$ for
    large system sizes. \label{fig:disEHS_Anderson3d}}
\end{figure}

\begin{figure*}
  \centering
  \begin{subfigure}{}%
    \includegraphics[width=0.31\textwidth]{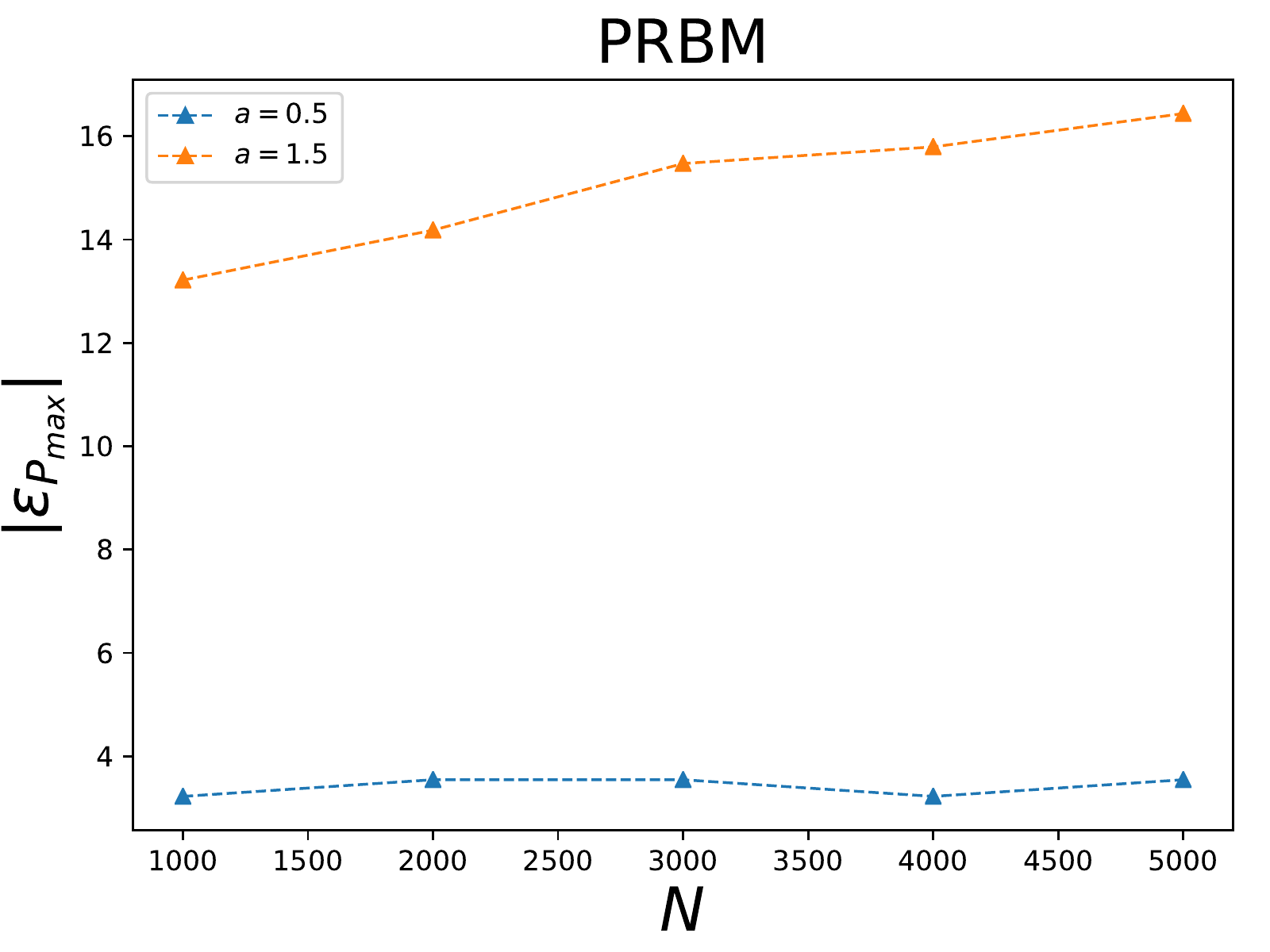}
  \end{subfigure}%
  ~%
  \begin{subfigure}{}%
    \includegraphics[width=0.31\textwidth]{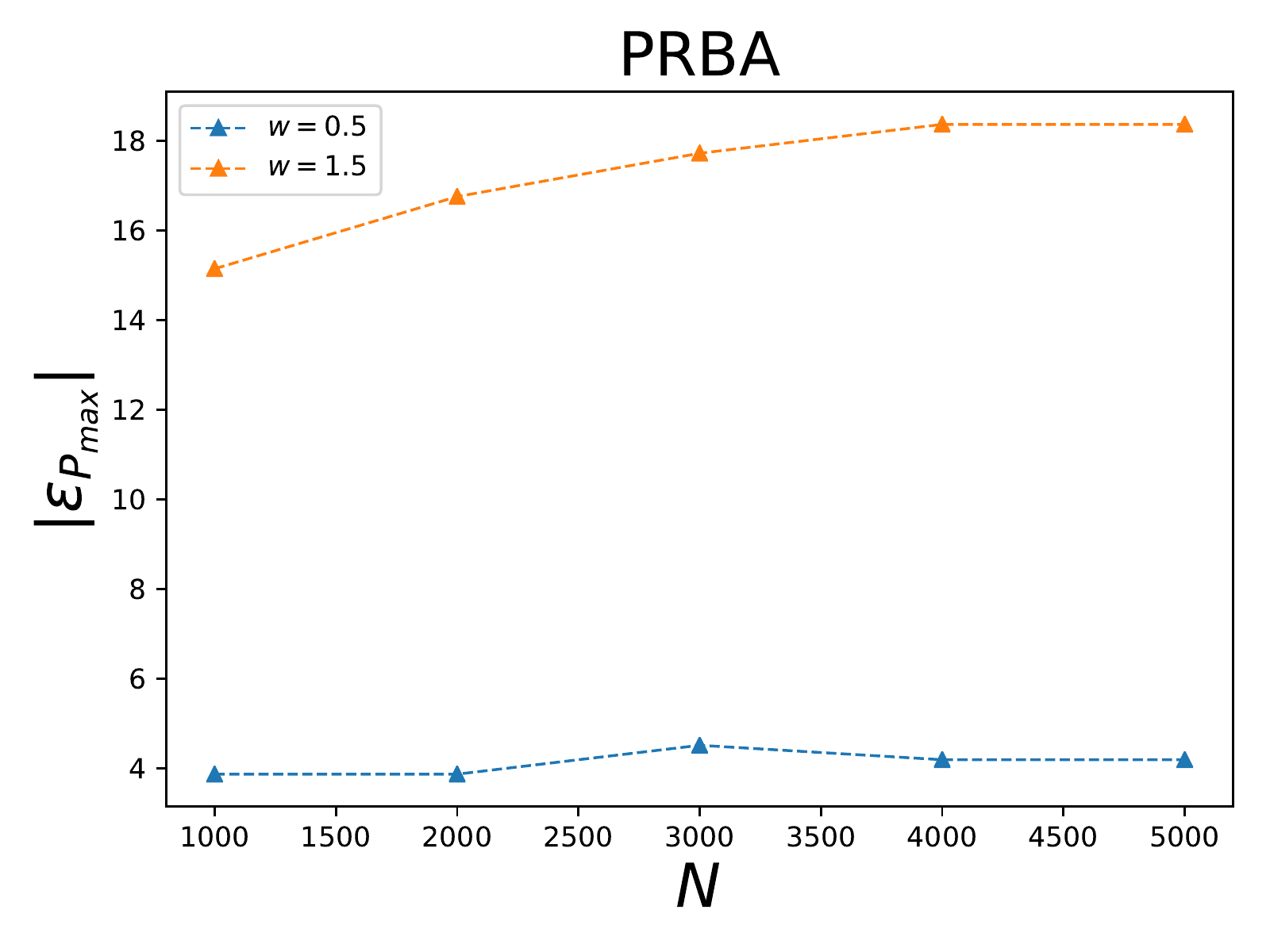}
  \end{subfigure}%
  ~%
  \begin{subfigure}{}%
    \includegraphics[width=0.31\textwidth]{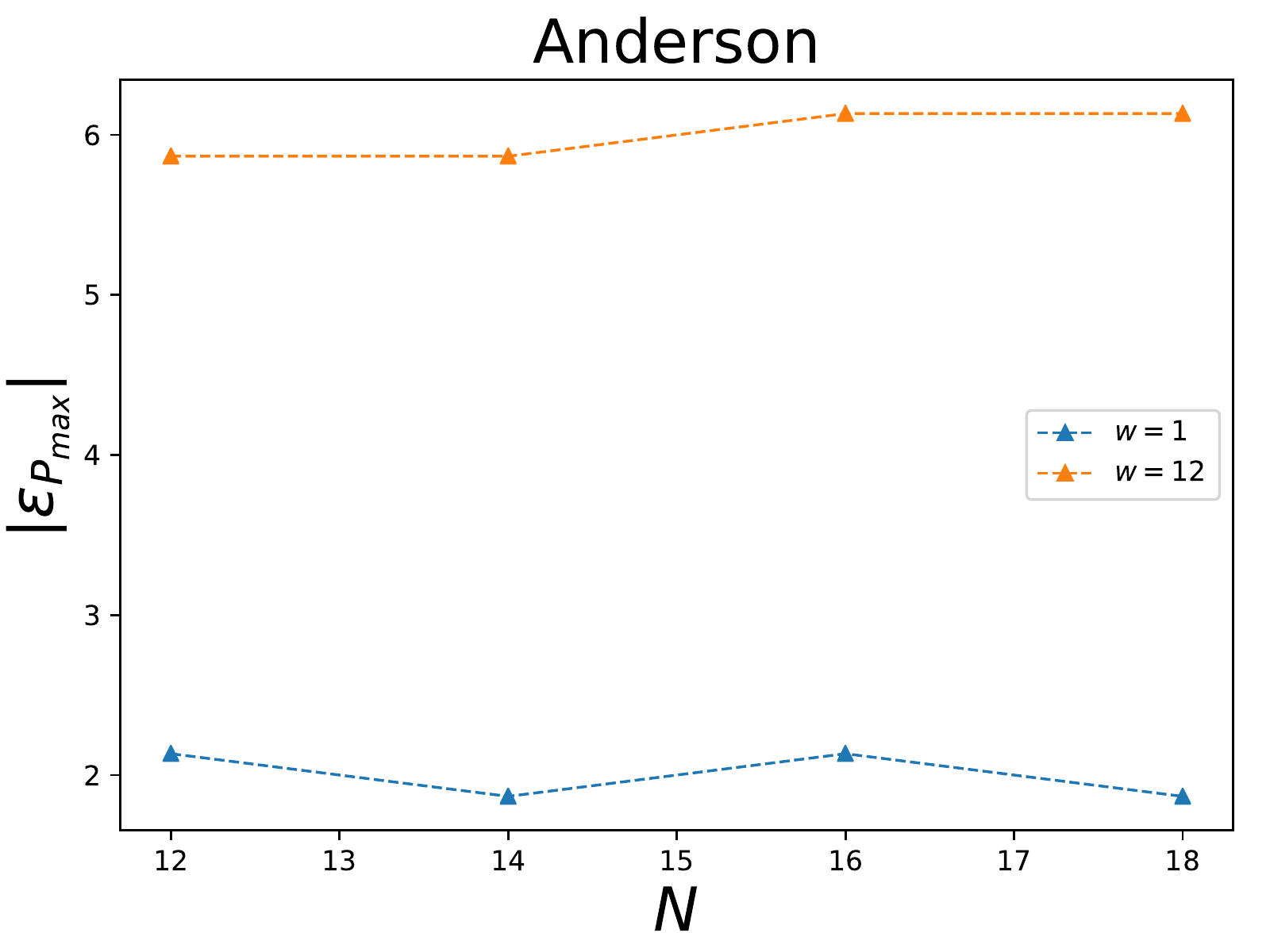}
  \end{subfigure}
  \caption{(color online) behaviour of the smallest magnitude
    entanglement Hamiltonian eigenvalue with largest probability
    $\epsilon_{\mathcal{P}_{max}}$ for delocalized and localized
    phases in PRBM (left panel), PRBA (middle panel), and Anderson
    $3d$ models (right panel). \label{fig:eps_pmax}}
\end{figure*}

In Figs.  \ref{fig:disEHS_PRBM}, \ref{fig:disEHS_PRBA}, and
\ref{fig:disEHS_Anderson3d} we plot distribution of the EHS for PRBM,
PRBA, and Anderson $3d$ models in delocalized and localized phases. We
see that for each system size $N$ in the localized phase, probability
distribution of $\epsilon$'s, $\mathcal{P}_{\epsilon}$, at small
magnitude $\epsilon$'s is negligible and the big share comes from
larger magnitude $\epsilon$'s, which according to Eq. (\ref{EE})
yields to low EE. In addition, as we increase system size, that
$\epsilon$ corresponds to maximum probability,
$\epsilon_{\mathcal{P}_{max}}$, shifts to larger magnitude (yielding
to smaller EE), and the corresponding probability increases (yielding
ro larger EE); i.e. two opposite factors causes EE to saturate. The
behaviour of the $\epsilon$ with highest probability
$\epsilon_{\mathcal{P}_{max}}$ in delocalized and localized phases as
we increase system size $N$ is plotted in Fig. \ref{fig:eps_pmax} for
three mentioned models.

On the other hand, distribution of EHS for delocalized phase is
noticeably different. Small values of $\epsilon$'s have big shares of
probability compare to large $\epsilon$'s; which yields to a large
EE. In addition, as we increase the system size,
$\epsilon_{\mathcal{P}_{max}}$ stays fixed (see
Fig. \ref{fig:eps_pmax}), but its probability increases. Thus, EE
becomes larger as we increase system size.

\section{Smallest magnitude entanglement Hamiltonian
  eigenvalue}\label{eps}
\begin{figure*}
  \centering
  \begin{subfigure}{}%
    \includegraphics[width=0.325\textwidth]{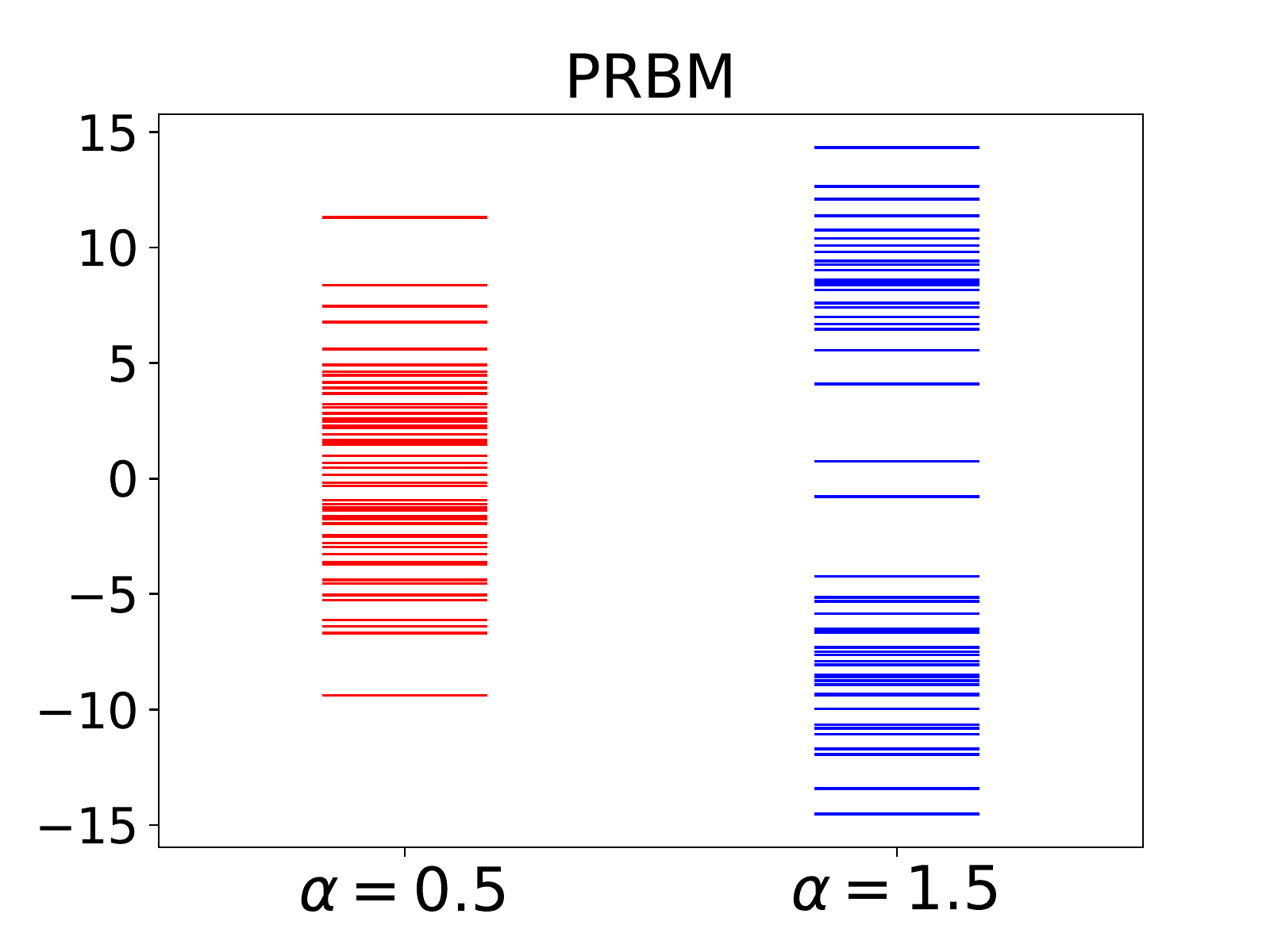}
  \end{subfigure}%
  ~%
  \begin{subfigure}{}%
    \includegraphics[width=0.325\textwidth]{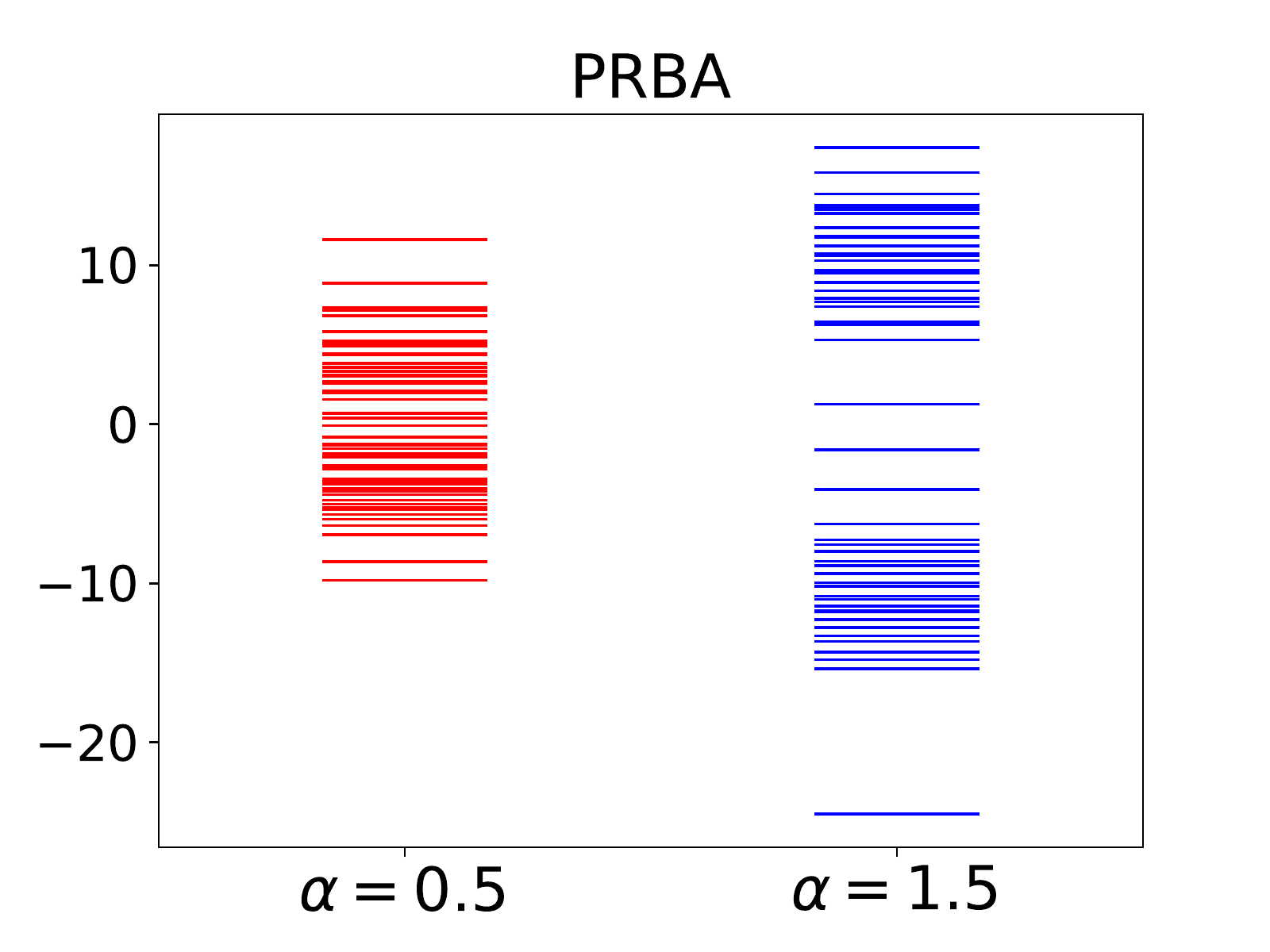}
  \end{subfigure}%
  ~%
  \begin{subfigure}{}%
    \includegraphics[width=0.325\textwidth]{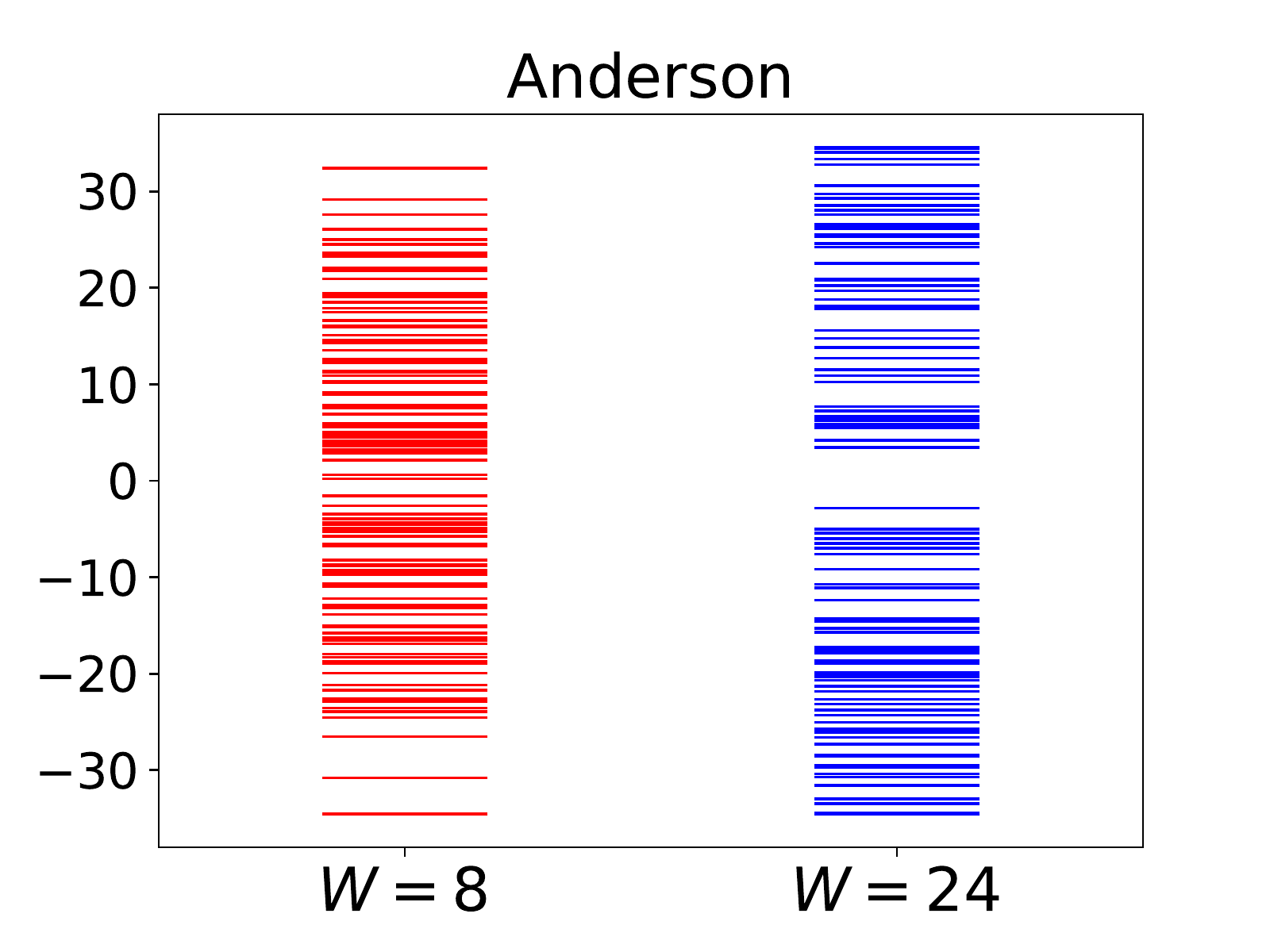}
  \end{subfigure}
  \caption{(color online) spectrum of entanglement Hamiltonian for
    PRBM (left panel), PRBA (middle panel), and Anderson $3d$ (right
    panel) models in delocalized (red) and delocalized (blue)
    phase. For PRBM, and PRBA system size $N=100$ and for Anderson
    $N=6\times 6 \times 6$. One sample is considered for each disorder
    strength without taking disorder average. \label{fig:spectrum}}
\end{figure*}

According to Figs.  \ref{fig:disEHS_PRBM}, \ref{fig:disEHS_PRBA} and
\ref{fig:disEHS_Anderson3d}, smallest magnitude $\epsilon$ has
distinguishable features in delocalized and localized phases. To see
it clearly, we plot the spectrum of EHS for one sample without taking
disorder average in Fig. \ref{fig:spectrum} for PRBM, PRBA, and
Anderson $3d$ model (Since Hamiltonian of the system has randomness,
we do not have particle-hole symmetry for eigenvalues of entanglement
Hamiltonian\cite{PhysRevB.69.075111}, so we do not expect a symmetric
distribution of the EHS). In delocalized phase, we have a (close to)
zero spectrum, while the smallest magnitude spectrum is a finite value
in the localized phase.

\begin{figure} \centering
  \begin{subfigure}{}%
\includegraphics[width=0.45\textwidth]{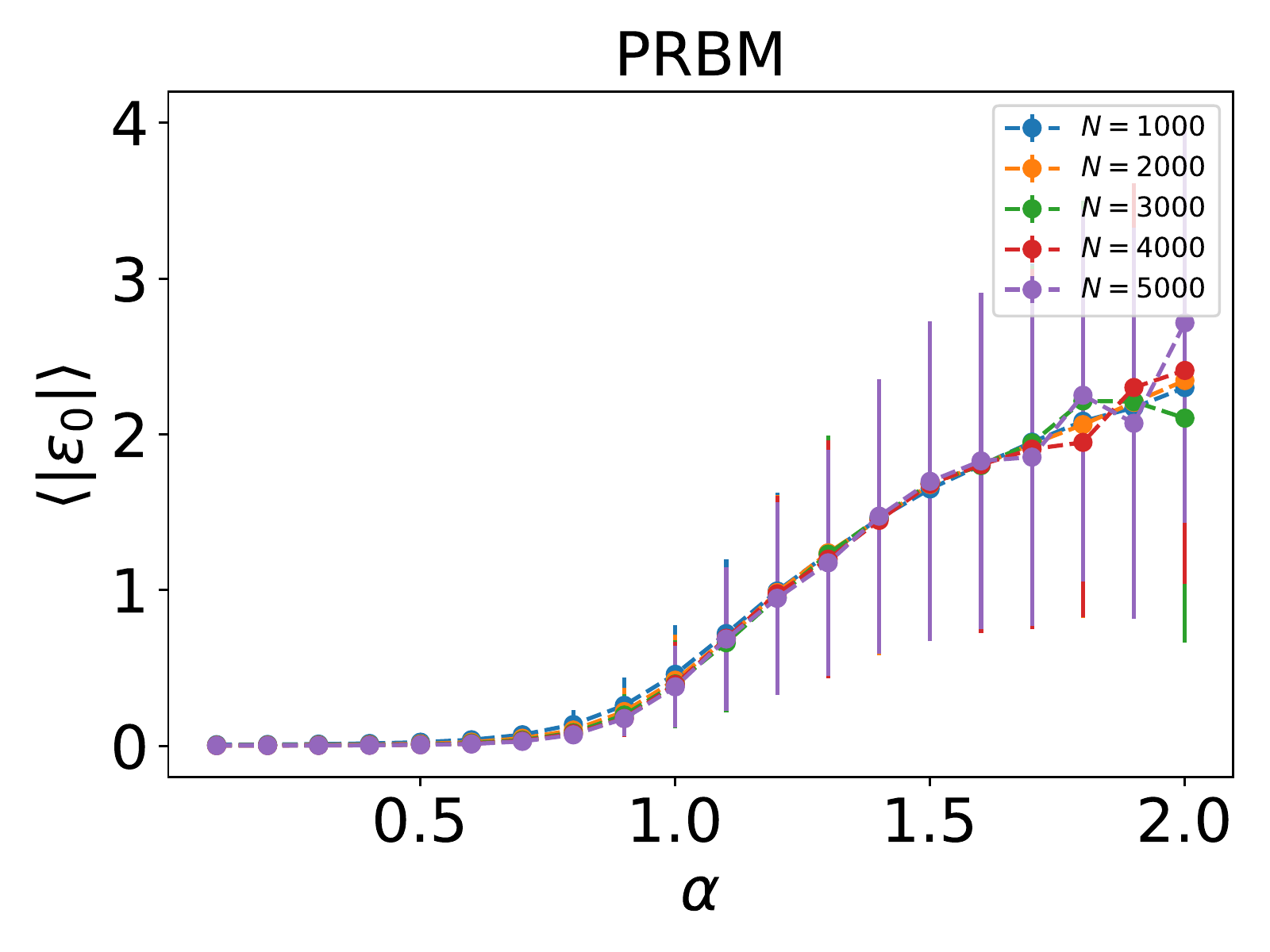}
  \end{subfigure}% ~%
  \begin{subfigure}{}%
\includegraphics[width=0.45\textwidth]{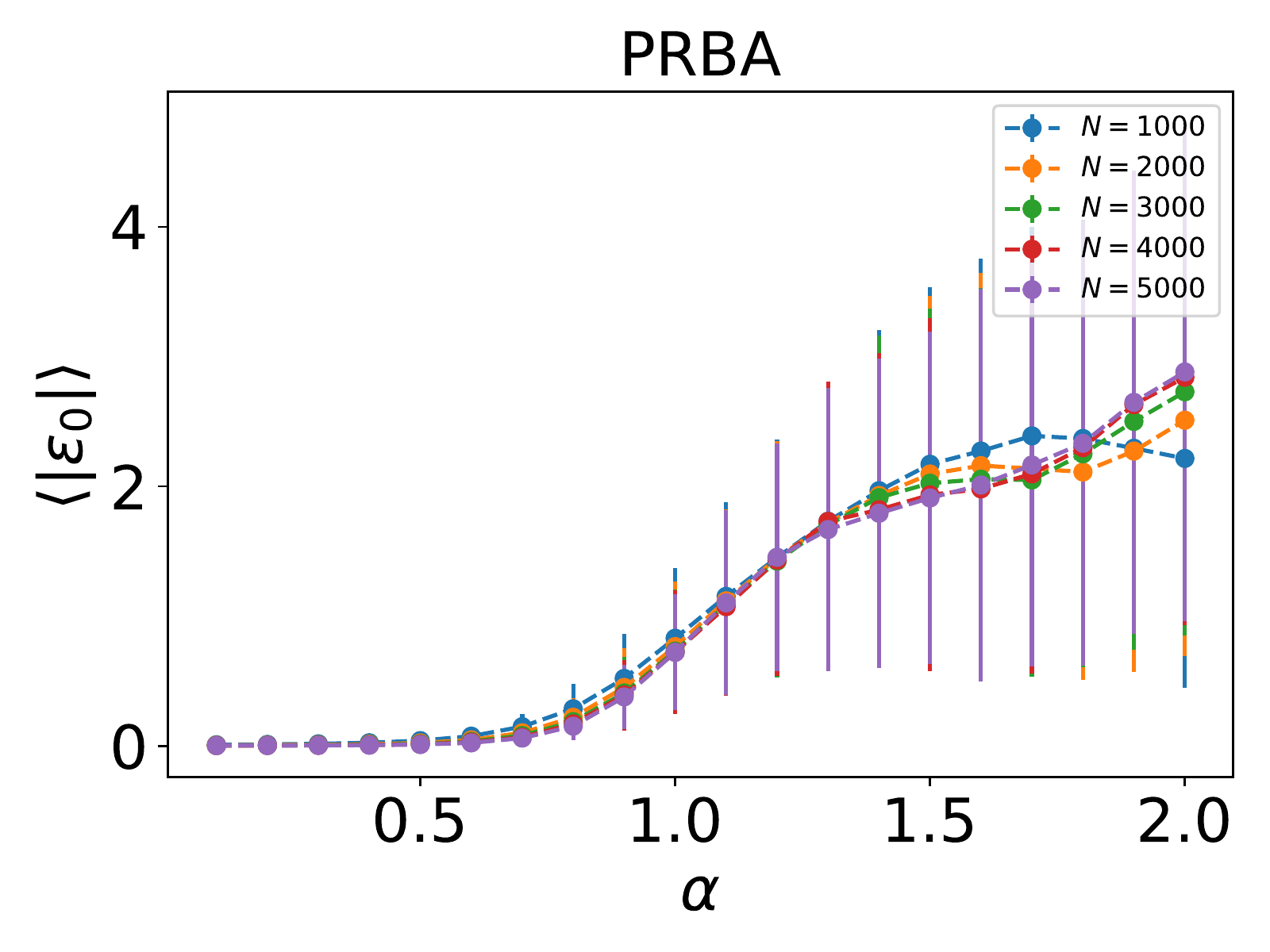}
  \end{subfigure}
  \caption{(color online) disorder averaged of smallest magnitude
    $\epsilon$ corresponding to non-zero probability distribution as
    we change disorder strength $a$ in PRBM model (upper panel), and
    PRBA model (lower model). In delocalized phase it is zero, and it
    moves toward larger values in localized phase. Vertical bar at
    each point shows the corresponding standard deviation.
    \label{fig:epsilon0_errorbar}}
\end{figure}

Accordingly, we propose smallest magnitude $\epsilon$ to be a
characterization of the delocalized-localized phase transition. We
plot disorder averaged smallest magnitude $\epsilon$ as we increase
disorder strength in Fig. \ref{fig:epsilon0_errorbar} for PRBM and
PRBA models. In delocalized phase the smallest magnitude $\epsilon$ is
zero, while it goes to larger $\epsilon$'s. We also note that
standard deviation of disorder averaged smallest $\epsilon$ is
considerably larger in the localized phase.

\section{Conclusion}\label{sum}
Entanglement in quantum system has been used vastly before for
characterizing phases and phase transitions in condensed matter
physics. EE diverges in delocalized phases and it saturates in
localized phase, thus behavior of the EE as we change the disorder
strength locates the phase transition point. In this report, by
employing free fermion models we explained that eigenvalues of the
entanglement Hamiltonian are also informative to characterize phases
and the phase transition point. In addition, we explained the behavior
of EE according to the distribution of the entanglement Hamiltonian
eigenvalues based on which we propose a characterization for the
delocalized-localized phase transition, namely the smallest magnitude
entanglement Hamiltonian eigenvalue. We applied this characterization
to our one dimensional models and found that its behavior is different
in delocalized and localized phases, although the phase transition
point is not sharply located by this characterization.

\acknowledgments
This work was supported by the University of Mazandaran.

\bibliographystyle{apsrev4-1.bst}
\bibliography{/home/cms/Dropbox/physics/Bib/reference.bib}
\end{document}